\documentclass[aps,showpacs,twocolumn]{revtex4}

\usepackage{amssymb}
\usepackage{amsmath}
\usepackage{bm}
\usepackage{graphics}
\usepackage{natbib}

\begin{document}
\title{Rashba-induced transverse pure spin currents in a four-terminal quantum dot ring}
\author{Weijiang Gong$^1$}
\author{Yu Han$^1$}
\author{Guozhu Wei$^{1,2}$}
\author{Yisong Zheng$^3$}\email[Author to whom correspondence should be
addressed. Email address: ]{zys@mail.jlu.edu.cn} \affiliation{1.
College of Sciences, Northeastern University, Shenyang 110004,
China\\ 2. International Center for Material Physics, Acadmia
Sinica, Shenyang 110015, China\\3. Department of physics, Jilin
University, Changchun 130023, China}

\date{\today}

\begin{abstract}
By applying a local Rashba spin-orbit interaction on an individual
quantum dot of a four-terminal four-quantum-dot ring and introducing
a finite bias between the longitudinal terminals, we theoretically
investigate the charge and spin currents in the transverse
terminals. It is found that when the quantum dot levels are separate
from the chemical potentials of the transverse terminals, notable
pure spin currents appear in the transverse terminals with the same
amplitude and opposite polarization directions. Besides, the
polarization directions of such pure spin currents can be inverted
by altering structure parameters, i.e., the magnetic flux, the bias
voltage, and the values of quantum dot levels with respect to the
chemical potentials of the transverse terminals.
\end{abstract}
\maketitle

\bigskip
Since the original proposal of the spin field effect transistor by
Datta and Das,\cite{Datta} enormous attention, from both
experimental and theoretical physics communities, has been devoted
to the controlling of the spin degree of freedom by means of the
spin-orbit (SO) coupling in the field of spintronics.\cite{Dasarma}
Particularly, in low-dimensional structures Rashba SO interaction
comes into play by introducing an electric potential to destroy the
symmetry of space inversion in an arbitrary spatial
direction.\cite{Rashba,Hirsch,Shytov,Sinova1,Das} Thus, based on the
properties of Rashba effect, electric control and manipulation of
the spin state is feasible. Accordingly, the Rashba-related
electronic properties in mesoscopic systems have been the main
concerns in spintronics, such as spin decoherence\cite{Loss,Souma}
and spin current\cite{Nagaosa,JH}.
\par
Quite recently, Rashba interaction has been introduced to coupled
quantum dot (QD) systems. Because the coupled QD systems possess
more tunable parameters to manipulate the electronic transport
behaviors, a number of interesting Rahsba-induced electron
properties are reported,\cite{Sun1,Chi} moreover, it is
theoretically predicted that pure spin currents are possibly
realized in a triple-terminal QD structures only by the presence of
a local Rashba interaction\cite{Guo,Gong-APL}. Following such a hot
topic, in this paper we propose a new theoretical scheme to realize
the pure spin current by virtue of Rashba interaction. We introduce
Rashba interaction to act locally on one component QD of a four-QD
ring with four terminals. Our theoretical investigation indicates
that the unpolarized charge current injected through the
longitudinal terminals gives rise to the emergence of pure spin
currents in transverse terminals with the same amplitude and
opposite polarization directions, and the polarization direction of
the pure spin current in either terminal is tightly dependent on the
adjustment of structure parameters.
\par
\begin{figure}
\centering \scalebox{0.43}{\includegraphics{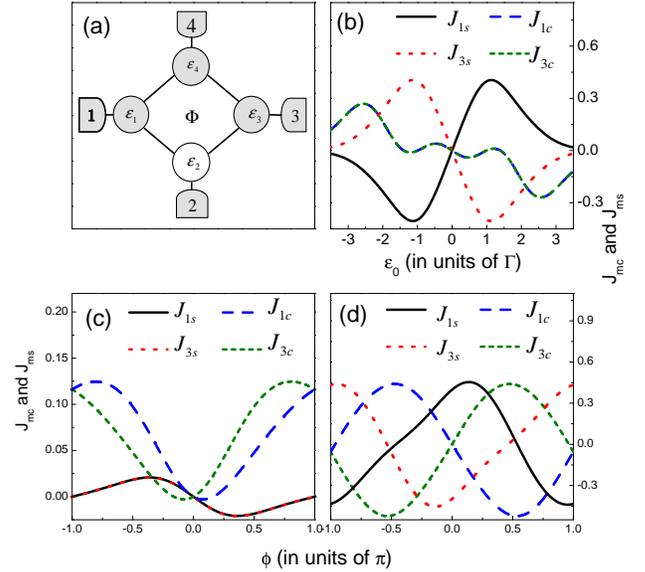}} \caption{ (a)
Schematic of a four-terminal four-QD ring structure with a local
Rashba interaction on QD-2. Four QDs and the leads coupling to them
are denoted as QD-j and lead-j with $j=1-4$. The currents vs the QD
levels $\varepsilon_j$ as well as the magnetic phase factor $\phi$
are shown in (b) and (c), respectively. The parameter values are
$\Gamma_j=2\Gamma$ and $\tilde{\alpha}=0.4$. In (b),
$\varepsilon_j=\varepsilon_0$ and $\phi=0$ and in (c), the QD level
$\varepsilon_j=\Gamma$. (d) The currents vs $\phi$ in the case
$\varepsilon_j=0$. \label{cond1}}
\end{figure}

The structure under consideration is illustrated in
Fig.\ref{cond1}(a). In such a system four leads are coupled to their
respective QDs in a four-QD ring, with the Rashba effect applied to
act locally on QD-2. The single-electron Hamiltonian in this
structure can be written as
$H_s=\frac{\textbf{P}^2}{2m^*}+V(\textbf{r})+\frac{\hat{y}}{2\hbar}\cdot[\alpha(\hat{\sigma}\times
\textbf{p})+(\hat{\sigma}\times \textbf{p})\alpha]$, where the
potential $V(\textbf{r})$ confines the electron to form the
structure geometry. The last term in $H_s$ denotes the local Rashba
SO coupling on QD-2. For the analysis of the electron properties, we
select the basis set
$\{\psi_{jk}\chi_{\sigma},\psi_j\chi_{\sigma}\}$($\psi_j$ and
$\psi_{jk}$ are the orbital eigenstates of the isolated QDs and
leads in the absence of Rashba interaction with $j$=1-4;
$\chi_{\sigma}$ with $\sigma=\uparrow,\downarrow$ denotes the
eigenstates of Pauli spin operator $\hat{\sigma}_z$) to
second-quantize the Hamiltonian, which is composed of three parts:
${\cal H}_{s}={\cal H}_{c}+{\cal H}_{d}+{\cal H}_{t}$.
\begin{eqnarray}
{\cal H}_{c}&&=\underset{\sigma jk}{\sum }\varepsilon
_{jk}c_{jk\sigma}^\dag c_{jk\sigma },\notag\\
{\cal H}_d&&=\sum_{j=1, \sigma}^{4}\varepsilon
_{j}d^\dag_{j\sigma}d_{j\sigma}
+\sum_{l=1,\sigma}^{2}[t_{l\sigma}d^\dag_{l\sigma}d_{l+1\sigma}+r_l(d_{l\downarrow}^\dag
d_{l+1\uparrow}\notag\\&&-d_{l+1\downarrow}^\dag
d_{l\uparrow})]+t_{3}d^\dag_{3\sigma}d_{4\sigma}+t_{4}e^{i\phi}d^\dag_{4\sigma}d_{1\sigma}+\mathrm
{H.c.},\notag\\ {\cal H}_{t}&&=\underset{\sigma jk }{\sum
}V_{j\sigma} d^\dag_{j\sigma}c_{jk\sigma}+\mathrm {H.c.},
\end{eqnarray}
where
$c_{jk\sigma}^\dag$ and $d^{\dag}_{j\sigma}$ $( c_{jk\sigma}$ and
$d_{j\sigma})$ are the creation (annihilation) operators
corresponding to the basis in lead-j and QD-j. $\varepsilon _{jk}$
and $\varepsilon_{j}$ are the single-particle levels.
$V_{j\sigma}=\langle
\psi_j\chi_\sigma|H|\psi_{jk}\chi_\sigma\rangle$ denotes QD-lead
coupling strength. The interdot hopping amplitude, written as
$t_{l\sigma}=t_l-i\sigma s_l$($l=1,2$), has two contributions:
$t_{l}=\langle\psi_{l}|H_0|\psi_{l+1}\rangle$ is the ordinary
transfer integral independent of the Rashba interaction;
$s_l=i\langle\psi_{l}|\alpha p_x+p_x\alpha|\psi_{l+1}\rangle$, a
real quantity for real $\psi_{l}$ and $\psi_{l+1}$, indicates the
strength of spin precession. Finally, in the Hamiltonian
$r_l=\langle\psi_{l}|\alpha p_z+p_z\alpha|\psi_{l+1}\rangle$ is a
complex quantity representing the strength of interdot spin flip. To
get an intuitive impression about the typical values of these
parameters in the Hamiltonian, we assume that each QD confines the
electron by an isotropic harmonic potential ${1\over
2}m^*\omega_0r^2$. Then, the four QDs distribute on a circle
equidistantly, and the interval in between is
$2l_0$($l_0=\sqrt{\hbar/m^*\omega_0}$). Besides, we assume that the
electron occupies the ground state in each QD. By defining a
dimensionless Rashba coefficient as
$\tilde{\alpha}=\alpha/(3\hbar\omega_0l_0)$, we obtain the rough
relation of the parameters: $t_1=t_2$, $s_1=s_2$, $r_1=-r_2$, and
$|s_l|=|r_l|\sim\tilde{\alpha} t_l$. Thereby we can express the
interdot hopping amplitude in an alternative form:
$t_{l\sigma}=t_l\sqrt{1+\tilde{\alpha}^2}e^{-i\sigma\varphi}$ with
$\varphi=\tan^{-1}\tilde{\alpha}$. Here the Rashba interaction
brings about a spin-dependent phase factor, which can be tuned by
varying the electric field strength. In the above Hamiltonian the
phase factor $\phi$ attached to $t_{4}$ accounts for the magnetic
flux through the ring. In addition, the many-body effect can be
readily incorporated into the above Hamiltonian by adding the
Hubbard term ${\cal
V}_{e\text{-}e}=\sum_{j\sigma}{\frac{U_j}{2}}n_{j\sigma}n_{j\bar{\sigma}}$.
\par
Starting from the second-quantized Hamiltonian, we can now formulate
the electronic transport properties. With the nonequilibrium Keldysh
Green function technique, the current flow in lead-$j$ can be
written as\cite{Meir}
\begin{equation}
J_{j\sigma}=\frac{e}{h}\sum_{j'\sigma'}\int d\omega
T_{j\sigma,j'\sigma'}(\omega)[f_j(\omega)-f_{j'}(\omega)],\label{current}
\end{equation}
where $T_{j\sigma,j'\sigma'}(\omega)=\Gamma_j
 G^r_{j\sigma,j'\sigma'}(\omega)\Gamma_{j'}G^a_{j'\sigma',j\sigma}(\omega)$
is the transmission function, describing electron tunneling ability
between lead-$j$ to lead-$j'$, and $f_j(\omega)$ is the Fermi
distribution function in lead-$j$. $\Gamma_j=2\pi
|V_{j\sigma}|^2\rho_j(\omega)$, the coupling strength between QD-j
and lead-j, can be usually regarded as a constant.  $G^r$ and $G^a$,
the retarded and advanced Green functions, obey the relationship
$[G^r]=[G^a]^\dag$. From the equation-of-motion method, the retarded
Green function can be obtained in a matrix form,
\begin{eqnarray}
&&[G^r]^{-1}=\notag\\
&&\left[\begin{array}{cccccccc} g_{1\uparrow}^{-1} & -t_{1\uparrow}&0&-t_{4}e^{-i\phi}&0&r^*_1&0&0\\
  -t^*_{1\uparrow}& g_{2\uparrow}^{-1}& -t_{2\uparrow}&0&-r^*_1&0&r^*_2&0\\
  0&-t^*_{2\uparrow}&g_{3\uparrow}^{-1}&-t_{3}&0&-r^*_2&0&0 \\
  -t_{4}e^{i\phi}&0&-t^*_{3}&g_{4\uparrow}^{-1}&0&0&0&0 \\
  0&-r_1&0&0&g_{1\downarrow}^{-1}&-t_{1\downarrow}&0&-t_{4}e^{-i\phi}\\
  r_1&0&-r_2&0&-t^*_{1\downarrow}& g_{2\downarrow}^{-1}& -t_{2\downarrow}&0\\
  0&r_2&0&0&0&-t^*_{2\downarrow}&g_{3\downarrow}^{-1}&-t_{3}\\
  0&0&0&0&-t_{4}e^{i\phi}&0&-t^*_{3}&g_{4\downarrow}^{-1}
\end{array}\right]\notag.
\end{eqnarray}
In the above expression, $g_{j\sigma}$ is the Green function of QD-j
unperturbed by the other QDs and in the absence of Rashba effect.
$g_{j\sigma}=[(z-\varepsilon_{j})\lambda_{j\sigma}+\frac{i}{2}\Gamma_j]^{-1}$
with $z=\omega+i0^+$ and
$\lambda_{j\sigma}=\frac{z-\varepsilon_{j}-U_{j}}{z-\varepsilon_{j}-U_{j}+U_j\langle
n_{j\bar{\sigma}}\rangle}$ resulting from the second-order
approximation of the Coulomb interaction\cite{Gongprb}.
$n_{j\sigma}=\langle d^\dag_{j\sigma}d_{j\sigma}\rangle$ can be
numerically resolved by iteration technique with $\langle
n_{j\sigma}\rangle=-\frac{i}{2\pi}\int d\omega
G^{<}_{j\sigma,j\sigma}$.

\par
We now proceed on to calculate the currents in lead-$j$, lead-1 and
lead-3 in this case. As for the chemical potentials in respective
leads, we consider $\mu_1$ as the zero point of energy of this
system and $\mu_1=\mu_3$; $\mu_2$ and $\mu_4$, the chemical
potentials in other two leads, are considered with
$\mu_2=\mu_1+eV/2$ and $\mu_4=\mu_1-eV/2$, in which $V$ is the bias
voltage. The charge and spin currents are defined respectively as
$J_{mc}=J_{m\uparrow}+J_{m\downarrow}$ and
$J_{ms}=J_{m\uparrow}-J_{m\downarrow}$ ($m=1,3$). Before
calculation, we introduce a parameter $\Gamma$ as the unit of
energy, the order of which is meV for some experiments based on
GaAs/GaAlAs QDs, as mentioned in the previous works\cite{PRL,PRB}.
\par
\begin{figure} \centering
\scalebox{0.45}{\includegraphics{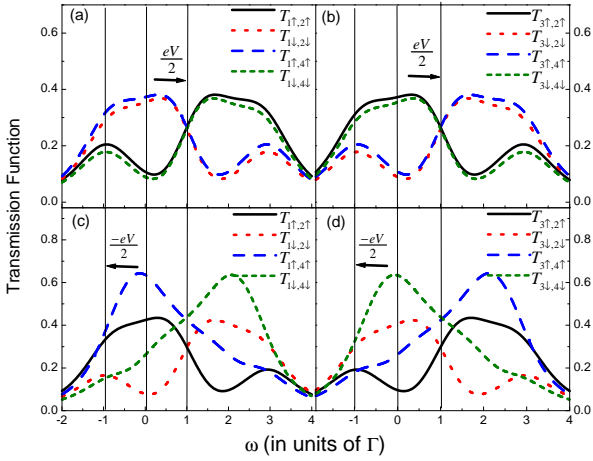}}\caption{ The spectra of
transmission functions $T_{m\sigma,m'\sigma}$($m$=1,3 and $m'$=2,4).
(a) and (b) Zero magnetic field case, and (c)-(d) magnetic phase
factor $\phi={\pi\over 2}$. The domains of integration to calculate
the currents are labeled by the arrows.\label{Trans}}
\end{figure}

To carry out the numerical calculation about the spectra of the
charge and spin currents in lead-1 and lead-3, we choose the Rashba
coefficient $\tilde{\alpha}=0.4$ which is available in the current
experiment.\cite{Sarra} The structure parameters are taken as
$|t_{l\sigma}|=t_3=t_4=\Gamma$, and the bias voltage is
$eV=2\Gamma$. In Fig.\ref{cond1}(b), the currents versus the QD
levels are shown in the absence of magnetic field. It shows that in
the case of the QD levels separate from the energy zero point (i.e.,
$\varepsilon_0\neq 0$), besides the emergence of pure spin currents,
a more interesting phenomenon is that the amplitude of $J_{1s}$ is
almost the same as that of $J_{3s}$. However, in the case of
$\varepsilon_0>0$ the value of $J_{1s}$ is greater than zero and
$J_{3s}<0$, whereas $J_{1s}<0$ and $J_{3s}>0$ under the condition of
$\varepsilon_0<0$. This means that, in such a structure applying a
finite bias between the longitudinal terminals can efficiently bring
out transverse pure spin currents with the opposite polarization
directions of them. Moreover, when the QD levels exceed the zero
point of energy of the system the polarization directions of these
pure spin currents will be thoroughly changed. However, for the case
of the QD levels consistent with the zero point of energy,
$\varepsilon_0=0$, no pure spin current comes about despite a
nonzero magnetic flux through the QD ring, as shown in
Fig.\ref{cond1}(c). When the QD levels take a finite value (eg.,
$\varepsilon_0=\Gamma$), as shown in Fig.\ref{cond1}(d), not only
there are apparent spin currents in the transverse terminals (lead-1
and lead-3), but also with the adjustment of magnetic flux in either
transverse terminal the charge and spin currents oscillate out of
phase. In the vicinity of $\phi =(n-{1\over2})\pi$, $J_{1c}$ and
$J_{3c}$ reach their maxima; Simultaneously, the spin current
$J_{ms}$ are just at a zero point. On the contrary, when $\phi
=n\pi$ the situation is just inverted, the maximum of $J_{ms}$
encounters the zero of $J_{mc}$. In particular, with the change of
magnetic phase factor from $\phi=2n\pi$ to $\phi=(2n+1)\pi$ the
polarization directions of the transverse pure spin currents are
inverted. So, it should be noticed that tuning the QD levels to an
nonzero value with respect to the zero point of energy is a key
condition of the appearance of transverse pure spin currents.
\par
The calculated transmission functions are plotted in Fig.
\ref{Trans} with $\varepsilon_j=\Gamma$. They are just the
integrands for the calculation of the charge and spin currents (see
Eq.(\ref{current})). By comparing the results shown in
Figs.\ref{Trans}(a) and \ref{Trans}(b), we can readily see that in
the absence of magnetic flux, the traces of
$T_{1\uparrow,2\uparrow}$, $T_{1\downarrow,4\downarrow}$,
$T_{3\downarrow,2\downarrow}$, and $T_{3\uparrow,4\uparrow}$
coincide with one another very well, so do the curves of
$T_{1\downarrow,2\downarrow}$, $T_{1\uparrow,4\uparrow}$,
$T_{3\uparrow,2\uparrow}$, and $T_{3\downarrow,4\downarrow}$.
Substituting such integrands into the current formulae, one can
certainly arrive at the result of the distinct pure spin currents,
which flows from lead-1 to lead-3 in such a case. On the other hand,
these transmission functions depend nontrivially on the magnetic
phase factor, as exhibited in Fig.\ref{Trans}(c) and (d) with
$\phi={\pi\over 2}$. In comparison with the zero magnetic field
case, herein the spectra of $T_{m\sigma,2\sigma}$ are reversed about
the axis $\omega=\Gamma$ without the change of their amplitudes, but
$T_{m\sigma,4\sigma}$ only present the enhancement of their
amplitudes. Similarly, with the help of Eq.(\ref{current}), one can
understand the disappearance of spin currents in such a case.
\par
The underlying physics being responsible for the spin dependence of
the transmission functions is quantum interference, which manifests
if we analyze the electron transmission process in the language of
Feynman path. Notice that the spin flip arising from the Rashba
interaction does not play a leading role in causing the appearance
of spin and charge currents\cite{Gong-APL}. Therefore, to keep the
argument simple, we drop the spin flip term for the analysis of
quantum interference. With this method, we write
$T_{1\sigma,2\sigma}=|\tau_{1\sigma,2\sigma}|^2$ where the
transmission probability amplitude is defined as
$\tau_{1\sigma,2\sigma}=\widetilde{V}^*_{1\sigma}
G^r_{1\sigma,2\sigma}\widetilde{V}_{2\sigma}$ with
$\widetilde{V}_{j\sigma}=V_{j\sigma}\sqrt{2\pi\rho_j(\omega)}$. By
solving $G^r_{1\sigma,2\sigma}$, we find that the transmission
probability amplitude $\tau_{1\sigma,2\sigma}$ can be divided into
three terms, i.e.,
$\tau_{1\sigma,2\sigma}=\tau^{(1)}_{1\sigma,2\sigma}+\tau^{(2)}_{1\sigma,2\sigma}
+\tau^{(3)}_{1\sigma,2\sigma}$, where
$\tau^{(1)}_{1\sigma,2\sigma}=\frac{1}{D}\widetilde{V}^*_{1\sigma}
g_{1\sigma}t_{1\sigma}g_{2\sigma}\widetilde{V}_{2\sigma}$, $
\tau^{(2)}_{1\sigma,2\sigma}=\frac{1}{D}\widetilde{V}^*_{1\sigma}
g_{1\sigma}t_{4}e^{-i\phi}g_{4\sigma}t^*_3g_{3\sigma}t^*_{2\sigma}
g_{2\sigma}\widetilde{V}_{2\sigma}$, and $
\tau^{(3)}_{1\sigma,2\sigma}=-\frac{1}{D}\widetilde{V}^*_{1\sigma}
g_{1\sigma}t_{1\sigma}g_{2\sigma}t_{2\sigma}g_{3\sigma}t^*_{2\sigma}g_{2\sigma}\widetilde{V}_{2\sigma}$
with $D=\det\{[G^r]^{-1}\}\prod_jg_{j\sigma}$. By observing the
structures of $\tau^{(1)}_{1\sigma,2\sigma} $,
$\tau^{(2)}_{1\sigma,2\sigma} $, and $\tau^{(3)}_{1\sigma,2\sigma}
$, we can readily find that they just represent the three paths from
lead-2 to lead-1 via the QD ring. The phase difference between
$\tau^{(1)}_{1\sigma,2\sigma}$ and $\tau^{(2)}_{1\sigma,2\sigma}$ is
$\Delta\phi^{(1)}_{2\sigma}=[\phi-2\sigma\varphi-\theta_3-\theta_4]$
with $\theta_j$ arising from $g_{j\sigma}$, whereas the phase
difference between $\tau^{(2)}_{1\sigma,2\sigma}$ and
$\tau^{(3)}_{1\sigma,2\sigma}$ is
$\Delta\phi^{(2)}_{2\sigma}=[\phi-2\sigma\varphi]$. It is clear that
only these two phase differences are related to the spin
polarization. $T_{1\sigma,4\sigma}$ can be analyzed in a similar
way. We then write
$T_{1\sigma,4\sigma}=|\tau^{(1)}_{1\sigma,4\sigma}+\tau^{(2)}_{1\sigma,4\sigma}+\tau^{(3)}_{1\sigma,4\sigma}|^2$,
with
$\tau^{(1)}_{1\sigma,4\sigma}=\frac{1}{D}\widetilde{V}^*_{1\sigma}g_{1\sigma}
t_{4}e^{-i\phi}g_{4\sigma}\widetilde{V}_{4\sigma}$,
$\tau^{(2)}_{1\sigma,4\sigma}=\frac{1}{D}\widetilde{V}^*_{1\sigma}g_{1\sigma}
t_{1\sigma}g_{2\sigma}t_{2\sigma}g_{3\sigma}t_3g_{4\sigma}\widetilde{V}_{4\sigma}$,
and
$\tau^{(3)}_{1\sigma,4\sigma}=-\frac{1}{D}\widetilde{V}^*_{1\sigma}
g_{1\sigma}t_4e^{-i\phi}g_{4\sigma}t^*_{3}g_{3\sigma}t_{3}g_{4\sigma}\widetilde{V}_{4\sigma}$.
The phase difference between $\tau^{(1)}_{1\sigma,4\sigma}$ and
$\tau^{(2)}_{1\sigma,4\sigma}$ is
$\Delta\phi^{(1)}_{4\sigma}=[\phi-2\sigma\varphi+\theta_2+\theta_3]$,
and $\Delta\phi^{(2)}_{4\sigma}=[\phi-2\sigma\varphi]$ originates
from the phase difference between $\tau^{(2)}_{1\sigma,4\sigma}$ and
$\tau^{(3)}_{1\sigma,4\sigma}$. Utilizing the parameter vlues in
Fig.\ref{Trans}, we evaluate that $\varphi\approx{\pi\over 7}$ and
$\theta_j=-\frac{3\pi}{4}$ at the point of $\omega=0$. It is
apparent that when $\phi=0$ only the phase differences
$\Delta\phi^{(1)}_{2\sigma}$ and $\Delta\phi^{(1)}_{4\sigma}$ are
spin-dependent. Accordingly, we obtain that
$\Delta\phi^{(1)}_{2\uparrow}=-\Delta\phi^{(1)}_{4\downarrow}={17\pi\over
14}$, and
$\Delta\phi^{(1)}_{2\downarrow}=-\Delta\phi^{(1)}_{4\uparrow}=-{3\pi\over
14}$, which clearly prove that the quantum interference between
$\tau^{(1)}_{1\uparrow,2\uparrow}$ and
$\tau^{(2)}_{1\uparrow,2\uparrow}$
($\tau^{(1)}_{1\downarrow,4\downarrow}$ and
$\tau^{(2)}_{1\downarrow,4\downarrow}$ alike) is destructive, but
the constructive quantum interference occurs between
$\tau^{(1)}_{1\downarrow,2\downarrow}$ and
$\tau^{(2)}_{1\downarrow,2\downarrow}$
($\tau^{(1)}_{1\uparrow,2\uparrow}$ and
$\tau^{(2)}_{1\uparrow,2\uparrow}$ alike). Then such a quantum
interference pattern can explain the traces of the transmission
functions shown in Fig.\ref{Trans}(a). In the case of
$\phi=\frac{1}{2}\pi$ we find that only
$\Delta\phi^{(2)}_{2(4)\sigma}$ are crucial for the occurrence of
spin polarization. By a calculation, we obtain
$\Delta\phi^{(2)}_{2(4)\uparrow}={5\pi\over 14}$ and
$\Delta\phi^{(2)}_{2(4)\downarrow}={9\pi\over 14}$, which are able
to help us clarify the results in Fig.\ref{Trans}(c) and (d). Up to
now, the characteristics of the transmission functions, as shown in
Fig.\ref{Trans}, hence, the tunability of charge and spin currents
have been clearly explained by analyzing the quantum interference
between the transmission paths.
\par
\begin{figure}[htb]
\centering \scalebox{0.45}{\includegraphics{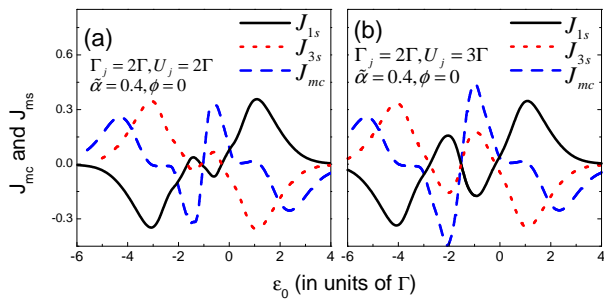}}\caption{ The
currents versus $\varepsilon_0$ in the presence of many-body terms
with $U_j=2\Gamma$ and $3\Gamma$, respectively.\label{Coulomb}}
\end{figure}
\par
So far we have not discuss the effect of electron interaction on the
occurrence of pure spin currents, though it is included in our
theoretical treatment. Now we incorporate the electron interaction
into the calculation, and we deal with the many-body terms by
employing the second-order approximation, since we are not
interested in the electron correlation here. Figure \ref{Coulomb}
shows the calculated currents spectra with $U_j=2\Gamma$ and
$3\Gamma$, respectively. Clearly, within such an approximation the
Rashba-induced transverse pure spin currents remain, though the
current spectra oscillate to a great extent with the shift of QD
levels.
\par
In conclusion, by introducing a local Rashba interaction on an
individual QD, we have studied the electronic transport through a
four-QD ring with four terminals. As a consequence, the
Rashba-induced transverse pure spin currents are observed by
applying a finite bias on the longitudinal terminals. The modulation
of the QD levels and the magnetic phase factor can efficiently
adjust the phases of the transmission paths, thus the spin-dependent
electron transmission probabilities can be controlled by tuning
these structure parameters, which brings out the change of the
amplitudes and directions of the pure spin currents. With respect to
the quantum interference in such a structure, we have to illustrate
two aspects. First, the applying of Rashba interaction is the
precondition of the spin-dependent electron transmission. The
presence of the multi-terminal configuration leads to the
comparative amplitudes of the different transmission paths for the
quantum interference. Finally, it should be emphasized that altering
the longitudinal bias, equivalent to interchange the sequence
numbers of lead-1 and lead-3, can also change the polarization
directions of the pure spin currents.

\clearpage

\bigskip


\begin{thebibliography}{99}
\bibitem{Datta} S. Datta and B. Das, Appl. Phys. Lett. \textbf{56}, 665 (1990).
\bibitem{Dasarma} S. A. Wolf, $et$ $al$., Science \textbf{294}, 1488 (2001);
Zutic, J. Fabian, and S. Das Sarma, Rev. Mod. Phys. \textbf{76}, 323
(2004).

\bibitem{Rashba} A. Bychkov and E. I. Rashba,
J. Phys. C \textbf{17}, 6039 (1984); J. Nitta, T. Akazaki, H.
Takayanagi, and T. Enoki, Phys. Rev. Lett. 78, 1335 (1997).
\bibitem{Hirsch} J. E. Hirsch, Phys. Rev. Lett. \textbf{83}, 1834 (1999).
\bibitem{Sinova1} J. Sinova, Phys. Rev. Lett. \textbf{92}, 126603 (2004).
\bibitem{Shytov} E. G. Mishchenko, A.V. Shytov, and B. I. Halperin, Phys. Rev.
Lett. \textbf{93}, 226602 (2004).
\bibitem{Das} I. \v{Z}uti\'{c}, J. Fabian, and S. Das Sarma, Appl. Phys. Lett.
\textbf{82}, 221 (2003); G. Schmidt, D. Ferrand, L. W. Molenkamp, A.
T. Filip, and vanWees B. J., Phys. Rev. B \textbf{62}, R4790 (2000).
\bibitem{Loss} D. V. Bulaev and D. Loss, Phys. Rev. B \textbf{71},
205324 (2005).

\bibitem{Souma} B. K. Nikolic and S. Souma, Phys. Rev. B \textbf{71},
195328 (2005).

\bibitem{Nagaosa} S. Murakami, N. Nagaosa, and S.-C. Zhang, Science \textbf{301},
1348 (2003).

\bibitem{JH} J. H. Bardarson, \.{I}. Adagideli, and Ph. Jacquod, Phys. Rev. Lett. \textbf{98}, 196601
(2007).




\bibitem{Sun1} Q. F. Sun, J. Wang, and H. Guo, Phys. Rev. B \textbf{71},
165310 (2005); Q.F. Sun and X. C. Xie, Phys. Rev. B \textbf{73},
235301 (2006).
\bibitem{Chi} F. Chi and S. Li, J. Appl. Phys. \textbf{100}, 113703 (2006).
\bibitem{Guo} H. F. L\"{u} and Y. Guo, Appl. Phys. Lett. \textbf{91},
092128(2007); F. Chi and J. Zheng, Appl. Phys. Lett. \textbf{92},
062106(2008).
\bibitem{Gong-APL} W. Gong, Y. Zheng, and
T. L\"{u}, Appl. Phys. Lett. \textbf{92}, 042104 (2008).
\bibitem{Meir} Y. Meir and N. S. Wingreen, Phys. Rev. Lett. \textbf{68}, 2512 (1992);
A.-P. Jauho, N. S. Wingreen, and Y. Meir, Phys. Rev. B \textbf{50},
5528 (1994).
\bibitem{Gongprb} J. Q. You and H. Z. Zheng, Phys. Rev. B 60, 13314 (1999); W. Gong, Y. Zheng,Y. Liu and
T. L\"{u}, Phys. Rev. B \textbf{73}, 245329 (2006).
\bibitem{PRL} M. Sigrist, T. Ihn, K. Ensslin, M. Reinwald, and W. Wegscheider,
Phys. Rev. Lett. \textbf{98}, 036805 (2007).
\bibitem{PRB} V. I. Puller and Y. Meir, Phys. Rev. B
\textbf{77}, 165421 (2008).
\bibitem{Sarra} J. Nitta, T. Akazaki, H. Takayanagi, and T. Enoki, Phys. Rev. Lett. \textbf{78},
1335 (1997); F. Mireles and G. Kirczenow, Phys. Rev. B \textbf{64},
024426 (2001); D. S\'{a}nchez and L. Serra, Phys. Rev. B
\textbf{74}, 153313 (2006).

\end{thebibliography}
\end{document}